\documentclass[prl,letterpaper,twocolumn]{revtex4}

\usepackage{graphicx}
\usepackage{dcolumn}
\usepackage{bm}

\begin{document} 

\title{Relativistic Cyclotron Radiation Detection of Tritium Decay Electrons as
  a New Technique for Measuring the Neutrino Mass} 
\author{Benjamin Monreal}
\affiliation{Department of Physics, University of California, Santa Barbara, CA}
\email{bmonreal@physics.ucsb.edu}
\author{Joseph A. Formaggio}
\affiliation{Laboratory for Nuclear Science and Department of Physics, Massachusetts Institute of Technology, Cambridge MA}

\date{\today}

\begin{abstract}The shape of the beta decay energy distribution is
  sensitive to the mass of the electron neutrino.  Attempts to measure
  the endpoint shape of tritium decay have so far seen no distortion from the zero-mass form, thus placing an upper
  limit of ${m_\nu}_\beta < $ 2.3 eV.   Here we show that a new type of
  electron energy spectroscopy could improve future
  measurements of this spectrum and therefore of the neutrino mass.
  We propose to detect the coherent cyclotron radiation emitted by an energetic electron in a magnetic field.  For
  mildly relativistic electrons, like those in tritium decay, the
  relativistic shift of the cyclotron frequency allows us to extract
  the electron energy from the emitted radiation.  We present calculations
  for the energy resolution, noise limits, high-rate measurement
  capability, and systematic errors expected in such an experiment.  
         \end{abstract}


\maketitle
\paragraph*{Introduction.}

Ever since Enrico Fermi's theory of beta decay \cite{fermi_tentativo_1933}, it has been known that
the neutrino mass has an effect on the decay kinematics. Measurements have always suggested that this mass is very small, with
successive experiments giving upper limits \cite{weinheimer_high_1999}\cite{lobashev_direct_1999}, most
recently ${m_\nu}_\beta < 2.3$ eV.  The upcoming KATRIN tritium
experiment\cite{angrik_katrin_2005} anticipates having a sensitivity of 0.20 eV at 90\% confidence.  Oscillation experiments, however,
tell us with great confidence that the tritium beta decay neutrinos are an admixture of at least two mass states, at least one of which has a nonzero mass, such that the effective mass must satisfy ${m_\nu}_\beta > 0.005$ eV under the normal hierarchy or  ${m_\nu}_\beta > 0.05$ eV in the inverted hierarchy \cite{farzan_effective_2002}.  The neutrino mass is an important component of precision cosmology \cite{Lesgourgues:2006nd}, and it may reflect physics at the GUT scale \cite{mohapatra_neutrino_1980}; this provides a strong motivation to find a way to measure tritium beta decay accurately enough to see ${m_\nu}_\beta$ down to the oscillation bounds. However, classical spectrometers are fundamentally limited by the need to transport electrons out of a source.

Tritium decays with a half-life of 12.32 y and maximum electron kinetic energy $E$ of
$E_0 = 18575$ eV; the effect of a nonzero neutrino mass is to shift this
maximum down to $E_0 - {m_\nu}_e$ and to suppress the phase
space within a few ${m_\nu}_e$ of this endpoint \cite{simkovic_exact_2007}.  We note two points
about the behavior of an 18575 eV electron in a magnetic field.  First, the electron
will follow a circular or spiral path with a cyclotron frequency of

\begin{equation}\label{eqw}
\omega = \frac{\omega_0}{\gamma} = \frac{qB}{m_e + E} 
\end{equation}

Note in particular that this
frequency depends on the electron Lorentz factor $\gamma$ and hence the electron kinetic energy $E$, but does not depend on
the pitch angle $\theta$, the angle between the electron velocity and the magnetic field direction.  Second, the electron emits
coherent cyclotron radiation \cite{johner_angular_1987} at frequency $\omega = 2\pi f$; for a wide range of parameters, the power emitted
is large enough to be detectable but not so large as to rapidly change the electron's energy.  This radiation spectrum therefore is sensitive to the electron energy, and its detection gives us a new form of non-destructive spectroscopy. 

\paragraph*{Experimental concept.}

\begin{figure}
\includegraphics[width=1.0\columnwidth]{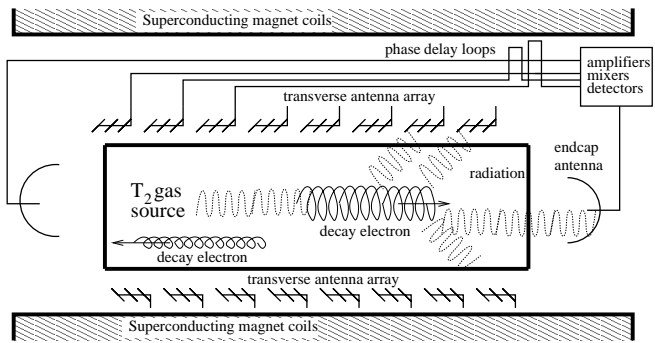}
\caption{Schematic of the proposed experiment.  A chamber encloses a diffuse gaseous tritium source under a uniform magnetic field.  Electrons produced from beta decay undergo cyclotron motion and emit cyclotron radiation, which is detected by an antenna array.  See text for more details.}
\label{diagram}
\end{figure}

Consider the arrangement shown in Fig. \ref{diagram}.  A low-pressure supply of
tritium gas is stored in a uniform magnetic field generated by a
solenoid magnet.  Tritium decay events release electrons with $0 < E < 18575$
eV (and velocity $0 < \beta < \beta_{e}$ where $\beta_e = 0.2625$) in random directions $\theta$ relative to the field vector.  The
electrons follow spiral paths with a velocity component 
$v_{||} = \beta cos(\theta)$ parallel to the magnetic field.  Each electron emits microwaves at
frequency $\omega$ and a total power which depends on $\beta$ and $\theta$
\begin{equation}
P(\beta,\theta) = \frac{1}{4\pi\epsilon_0}\frac{2q^2\omega_0^2}{3c}\frac{\beta^2 sin^2(\theta)}{1-\beta^2}
\end{equation}
 which are detected by an antenna array.  We propose to detect the radiation and measure its frequency spectrum, thus obtaining $\omega$ and hence $E$.

Although the emitted radiation is narrowband with frequency $\omega$, the signal seen in a stationary antenna is more complicated; generally it includes a Doppler shift due to $v_{||}$, some dependence on the electron-antenna distance (generally r$^{-2}$, but possibly complicated by reflections), and the differential angular power distribution of the emission.  The detected signal thus depends on the antenna configuration, and may have a nontrivial frequency content.   We discuss two candidate antenna choices and the signal expected in each.  

In the first case, we place the tritium source inside of a waveguide, and collect the microwaves with two ``endcap'' antennae at the ends of the tube.  For each electron, both antennae will see Doppler-shifted radiation (one redshift, one blueshift) due to the motion of the guiding center.  If we can detect both of these components, both the electron energy and pitch angle are uniquely determined.  

\begin{figure}
\includegraphics[width=0.95\columnwidth]{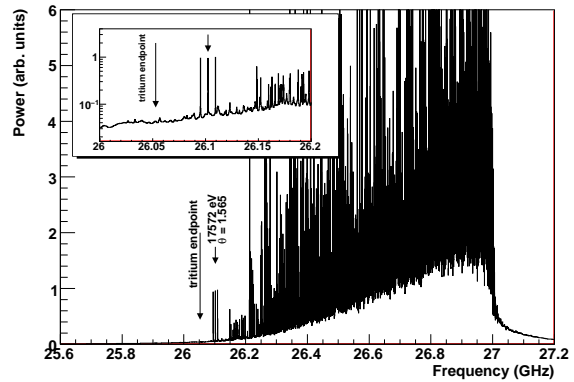}
\caption{Simulated microwave spectrum, showing the cyclotron emission of 10$^5$ tritium decays over 30$\mu$s in a 10m long uniform magnet ($\omega_0/2\pi$ = 27 GHz, B $\sim$ 1 T) with a finely-spaced transverse antenna array.   e$^-$-T$_2$ scattering is neglected.  The short arrow points out a triplet of spectral peaks generated by an individual high-energy, high-pitch angle electron; the central peak is the cyclotron frequency and the sidebands are due to AM modulation.  The log-scale inset zooms in on this electron and the endpoint region.}\label{fig2}         
\label{spectrum}
\end{figure}

In the second case, consider a long array of evenly-spaced antennae oriented transverse to the magnetic field.  Any single transverse antenna may see the electron passing by, resulting in a complex, broadband ``siren'' signal which sweeps from blueshift to redshift.  However, the coherent sum signal from all of the antennae in the array must be quasi-periodic.  If the antennae are spaced closely enough, and their signals summed with an appropriate choice of delay lines, almost all of the complex Doppler effects sum incoherently across the array, while the unshifted cyclotron frequency sums coherently.  The final summed periodic signal appears as a ``carrier wave'' at frequency $\omega$ with (a) an amplitude modulation, because the antenna response varies periodically along the electron's path, and (b) possibly a small residual frequency modulation due to the relativistic ``beaming'' of the cyclotron radiation (see Figure~\ref{spectrum}).  In frequency space, these modulations appear as sharp sidebands of the cyclotron frequency. 

\paragraph*{Energy resolution.}

In order to measure the electron energy to a precision $\Delta E$,
we need to measure the frequency to a relative precision of $\Delta f/f = \Delta E/m_e$.
For $\Delta E = 1$ eV this implies $\Delta f/f = 2\times 10^{-6}$.  In order to achieve a frequency precision of $\Delta f$, we need to monitor the signal for $t_{min} = 2/\Delta f$, according to Nyquist's theorem. This is a key number for several aspects of the experiment; for concreteness,
we discuss a reference design with a 1T magnetic field and a $\Delta E=1.0$ eV energy resolution.  First, we want the beta electrons to have mean free flight times longer than $t_{min}$ (30$\mu s$ in the reference design).   Due to T$_2$-e$^-$ scattering, this places a constraint on the density of the source.   The T$_2$-e$^-$ inelastic scattering cross section \cite{aseev_energy_2000} at 18 keV is $\sigma_i = 3\times10^{-18}$ cm$^2$, so in order to achieve the appropriate mean free path the T$_2$ density cannot exceed $\rho_{max} = (t_{min} \beta c \sigma)^{-1}$ ($1.4\times10^{11}$/cm$^3$ or 4 $\mu$Torr in the reference design).   It also places a constraint on the physical size of the apparatus; we presume that our measurement ends when the particle reaches the end of some instrumented region, although this is not necessarily the case.  If we want to be able to measure particles with minimum pitch angle $\theta_{min}$, the instrumented region needs to be of length $l = t_{min} \beta c \cdot cos(\theta_{min})$ long; in practice, engineering constraints on $l$ may set $\theta_{min}$.   Finally, $t_{min}$ also places a constraint on the magnetic field.  The electron continuously loses energy via cyclotron radiation; we want to complete our frequency measurement before it has lost energy $\Delta E$ due to radiative emission.   

\paragraph*{Bandwidth and data rate.}

One great advantage of the MAC-E filter technique used by experiments such as Mainz \cite{weinheimer_high_1999}, Troitsk, \cite{lobashev_direct_1999} and
KATRIN is the ability to effortlessly reject extremely
large fluxes of low-energy electrons, and to activate the detector and DAQ
only for the small fraction of decays near the endpoint.   A cyclotron
emission spectrometer will be exposed to all of the tritium decays in
its field of view (Fig. \ref{fig2}); therefore, it is important that we be able to process these decays without unreasonable pileup.

The main tool for separating signal from background is the
high-resolution and high-linearity nature of frequency domain
analysis.  Electrons with $E=0$ will generate fundamental signals at $f =$ 27.992490 GHz; 18.575 keV electrons will emit fundamentals at about 27.010643 GHz; as each 1 eV analysis bin is about 50 kHz wide the full region-of-interest (ROI) is perhaps 1 MHz wide.  Detecting a narrow signal in the endpoint ROI is, by itself, insufficient to confidently identify an endpoint electron, since this band is also populated by the low-frequency sidebands of the much more numerous low-energy electrons; we will need to detect at least two spectral lines, possibly three, in order to confidently identify an electron.   Any possible confusion source has a lower \emph{power} than a real ROI source at the same frequency, but because power measurements will be noisy, to be conservative we choose not to rely on them.

Several other parameters conspire to reduce the impact of sideband confusion.  In order for a low-energy electron to put any sideband at all into the ROI, it must have a large $v_{||}$ to generate the Doppler shift; however, a large $v_{||}$ also leads to a quick exit from the spectrometer (and consequently a broad signal) and to lower emitted power (both due to the quick exit and the $\beta_\perp^2$ term in the power).  Also, the inelastic scattering cross section increases like $1/E$ for low-energy electrons; if our source is filled with gas such that endpoint electrons have only the minimum tolerable path length, then low-energy electrons will be suppressed (or their signals collisionally broadened) by a large factor.   Accidental coincidences may still occur, however.  If the detection criterion requires simply two high-power spectral peaks in coincidence, we estimate that a T$_2$ source strength of 10,000 Bq would give an accidental-trigger rate comparable to KATRIN's background event rate of one per $10^{13}$ effective source decays.  Requiring a third spectral peak raises this allowable source strength to approximately $10^{9}$ Bq.  

In practice, we expect to have many cross-checks to remove accidental coincidences, such as location and polarization data, phase and power relationships between peaks, and so on.  Furthermore, for a real multiplet of peaks, all components will appear and disappear from the spectrum at the same time; we expect to be able to suppress the final accidental coincidence rate by at least a few orders of magnitude.  Full development of these analyses is beyond the scope of this paper.

The data acquisition rate of our system will be determined by the \emph{full} bandwidth over which we will search for spectral peaks, including fundamentals and sidebands.  This could be 100 MHz in the transverse-antenna case or several GHz in the endcap antenna case.

\paragraph*{Power and noise.}

It is important that single electrons can be detected well above the noise level; first of all, to avoid false events from noise fluctuations; secondly, in order to approach as nearly as possible the Nyquist limit on the frequency resolution; thirdly, to increase the precision of total-power measurements and start/stop time estimates for each detected electron.  

For our reference design with B=1 T, $\Delta E = 1 eV$, each resolution bin covers 50 kHz.  This bandwidth shows a thermal noise power of $6.5 \times 10^{-19}$ W/K, compared with a possible signal power in the neighborhood of $10^{-15}$ W.  In this frequency band, widely available amplifiers have 10-20K noise temperatures.  For lower magnetic fields, the signal strength varies as B$^2$ while the endpoint bandwidth varies as B, so the signal to noise gets worse.  

A second noise source comes from the incoherent signals of non-endpoint and/or low-pitch beta electrons.  For our 1-T, 30 $\mu$s-analysis-period reference design, each 50 kHz analysis bin near 26 GHz will show approximately 10$^{-24}$ W/Bq of tritium noise.  This is compatible with robust signal detection in the presence of the $10^8$--$10^9$ Bq source allowed by pileup limitations.  We note that this nonthermal power will have non-Gaussian fluctuations.




\paragraph*{Systematic errors.} 

This technique presents a very different systematic error budget than MAC-E filter experiments.  The spectrometer is continuously monitoring all decay energies, and thus is immune to slow source strength drifts.  We anticipate using an essentially static tritium gas whose electrostatic potential is fixed at both ends; this precludes large systematics due to source charging, voltage supply stability, flow-related Doppler shifts, and T$^-$ ion traps.  Microwave frequency measurements are easily stabilized against drifts at the $10^{-12}$ level.

We have two defenses against magnetic field drifts; first, NMR probes can monitor the total field to a precision of $10^{-7}$ or better. Secondly, a weak $^{83m}$Kr conversion electron source \cite{dragoun_increased_2004} could be injected directly into the source region, and the position and width of its narrow 30.5 keV L$_3$-32 line could be monitored with very high precision.  This would also monitor the mechanical stability of the detector array, field direction shifts, and data-analysis biases.  Magnetic field inhomogeneities are a source of concern; from the perspective of a purely frequency-domain analysis, they would be a source of line broadening.  High-precision energy analysis in a realistic instrument may need to rely on mixed time and frequency domain analysis, or even pure time-domain pulse fitting.   Since these complications are ``lossless'', we suppose for now that we will eventually be able to recover near-ideal frequency precision in spite of small inhomogeneities.

With respect to scattering, the situation of a microwave spectrometer is unusual.  This spectrometer has the ability to run with very low source column densities, and therefore to avoid large spectral distortion due to e$^-$-T$_2$ scattering, and the attendant uncertainties.  On the other hand, we have seen that a larger source column density may be useful for rapidly scattering low-energy electrons and preventing them from generating narrow-line signals, but this may re-insert the scattering systematic error.  An additional scattering-related uncertainty arises in this instrument: unlike a MAC-E spectrometer, the microwave instrument is not aperture-free.  Electrons can interact with the walls of the source tube and remain visible to the antenna array. (A related problem is that of high-energy electrons produced in the wall by radioactive sources.)  There are two possible avenues for avoiding scattering-related systematics.  First, we could use multi-antenna measurements to fiducialize a surface-free, apertureless source.  Also, a real scattering event does not simply \emph{change} the electron's energy, but it also broadens or splits its cyclotron emission line; we may be able to detect such scattering on an event-by-event basis.  

The excitation spectrum of (T$^3$He)$^+$ daughter ion \cite{Jonsell:1999aq} is unaffected by any improved T$_2$ measurement technique; it imposes an irreducible 0.36 eV energy spread on high-energy decays, with a difficult-to-estimate systematic error which the KATRIN collaboration estimates as contributing $\Delta m^2 < 6\times10^{-3}$ eV$^2$.  It is worth considering whether a future experiment of this type could use an atomic $T$ source; this possibility is beyond the scope of this paper.  

\paragraph*{Neutrino mass sensitivity.}

For simplicity, consider an T$_2$ experiment which merely counts the T$_2$ decay rate into the range $E_0 - 1.0 eV < E < E_0$ with no background.  This quantity is $3.5 \times 10^{-13}$ in the zero-mass neutrino case and $3.3\times 10^{-13}$ for an effective neutrino mass of $m_\nu = 0.2$ eV.  In the absence of systematic errors, in order to distinguish these cases with 95\% confidence, the experiment would need to observe a total of $3\times10^{15}$ tritium decays, or about $10^8$ Bq-years, which is comparable to our rough estimate of a single confusion-limited data channel.  Sensitivity to $m_v = 0.1 eV$, comparable with a KATRIN-like systematic error of $\Delta m^2 = 0.01$ eV$^2$ would require only $2 \times 10^9$ Bq y, which could be achieved by using multiple antenna arrays simultaneously to suppress pileup.

To increase the energy resolution of a microwave spectrometer, we require longer and longer observation times.  Surprisingly, this tends to \emph{improve} the single-event sensitivity of the experiment, since the long integrations and narrow bandwidths give us addition noise suppression capabilities.   We see no fundamental barrier to improving the energy resolution to 0.36 eV, the irreducible width due to final state excitations.  A more detailed estimate will require a better-specified experimental model with realistic noise, scattering systematics, and signal extraction.

Even in the absence of additional details, we wish note the parameters of an experiment with sensitivity to a neutrino mass as low as 0.007 eV, at the mass scale suggested by solar and reactor neutrino oscillation data.   First, the situation unambiguously demands an atomic tritium source, with extremely small molecular tritium contamination; such a source has yet to be developed.  Suppose we demand an energy resolution of 0.03 eV; in light of radiative broadening this demands a magnetic field of about B = 0.01 T and observations lasting 100 ms to obtain this precision on 270 MHz cyclotron radiation.  At this low frequency, although the emitted power is much reduced, the narrower bandwidth contributes to an acceptable signal-to-noise ratio.  In the absence of systematic errors, a 2$\sigma$ detection of an 0.007 eV neutrino mass requires $10^{20}$ events, or $4\times10^{12}$ Bq-y (100 Ci-y) of decay data; since we have specified a 100 ms dwell time per observation, a 100 Ci source would supply $3.7\times 10^{11}$ electrons per time bin.  Our previous pileup limit of $\sim 3000$ decays per time bin will be relaxed by a factor of the frequency resolution \footnote{The actual rate of accidental triplets will decrease with $\Delta f^3$, but the background rate requirement grows stricter as $\Delta f^{-2}$.}, so we can run the experiment with no more than $10^5$ decays per time bin.  (While this is an optimistic background- and systematics-free picture of the statistical situation, this treatment of the pileup remains fairly conservative.)  Under these circumstances, we can accumulate the desired statistics with $3\times10^6$ parallel measurement channels.  

In summary, we show that it is possible to measure the energies of tritium beta-decay electrons by spectroscopy of their cyclotron radiation emission in a magnetic field.  With this technique we can perform an array of new and powerful measurements of the endpoint of tritium beta decay.  

\paragraph*{Acknowledgments.}
The authors wish to thank Peter Doe, Shep Doeleman, Asher Kaboth, Michelle Leber, Phil Lubin, Michael Miller, Hamish Robertson, Leslie Rosenberg, and Gray Rybka for much useful input.  JF is supported by the Department of Energy Office of Nuclear Physics, Grant DE\-FG02\-06ER41420.

\bibliographystyle{apsrev}
\bibliography{project8}

\begin{thebibliography}{12}
\expandafter\ifx\csname natexlab\endcsname\relax\def\natexlab#1{#1}\fi
\expandafter\ifx\csname bibnamefont\endcsname\relax
  \def\bibnamefont#1{#1}\fi
\expandafter\ifx\csname bibfnamefont\endcsname\relax
  \def\bibfnamefont#1{#1}\fi
\expandafter\ifx\csname citenamefont\endcsname\relax
  \def\citenamefont#1{#1}\fi
\expandafter\ifx\csname url\endcsname\relax
  \def\url#1{\texttt{#1}}\fi
\expandafter\ifx\csname urlprefix\endcsname\relax\def\urlprefix{URL }\fi
\providecommand{\bibinfo}[2]{#2}
\providecommand{\eprint}[2][]{\url{#2}}

\bibitem[{\citenamefont{Fermi}(1933)}]{fermi_tentativo_1933}
\bibinfo{author}{\bibfnamefont{E.}~\bibnamefont{Fermi}},
  \bibinfo{journal}{Ricerca Scient.} \textbf{\bibinfo{volume}{2}},
  \bibinfo{pages}{12} (\bibinfo{year}{1933}).

\bibitem[{\citenamefont{Weinheimer et~al.}(1999)}]{weinheimer_high_1999}
\bibinfo{author}{\bibfnamefont{C.}~\bibnamefont{Weinheimer}}
  \bibnamefont{et~al.}, \bibinfo{journal}{Phys. Lett.}
  \textbf{\bibinfo{volume}{B460}}, \bibinfo{pages}{219} (\bibinfo{year}{1999}).

\bibitem[{\citenamefont{Lobashev et~al.}(1999)}]{lobashev_direct_1999}
\bibinfo{author}{\bibfnamefont{V.~M.} \bibnamefont{Lobashev}}
  \bibnamefont{et~al.}, \bibinfo{journal}{Phys. Lett.}
  \textbf{\bibinfo{volume}{B460}}, \bibinfo{pages}{227} (\bibinfo{year}{1999}).

\bibitem[{\citenamefont{Angrik et~al.}(2005)}]{angrik_katrin_2005}
\bibinfo{author}{\bibfnamefont{J.}~\bibnamefont{Angrik}} \bibnamefont{et~al.}
  (\bibinfo{year}{2005}), \bibinfo{note}{{FZKA-7090}},
  \urlprefix\url{http://bibliothek.fzk.de/zb/abstracts/7090.htm}.

\bibitem[{\citenamefont{Farzan and Smirnov}(2002)}]{farzan_effective_2002}
\bibinfo{author}{\bibfnamefont{Y.}~\bibnamefont{Farzan}} \bibnamefont{and}
  \bibinfo{author}{\bibfnamefont{A.~Y.} \bibnamefont{Smirnov}},
  \bibinfo{journal}{hep-ph/0211341}  (\bibinfo{year}{2002}),
  \bibinfo{note}{{Phys.Lett.} B557 (2003) 224-232}.

\bibitem[{\citenamefont{Lesgourgues and Pastor}(2006)}]{Lesgourgues:2006nd}
\bibinfo{author}{\bibfnamefont{J.}~\bibnamefont{Lesgourgues}} \bibnamefont{and}
  \bibinfo{author}{\bibfnamefont{S.}~\bibnamefont{Pastor}},
  \bibinfo{journal}{Phys. Rept.} \textbf{\bibinfo{volume}{429}},
  \bibinfo{pages}{307} (\bibinfo{year}{2006}), \eprint{astro-ph/0603494}.

\bibitem[{\citenamefont{Mohapatra and
  Senjanović}(1980)}]{mohapatra_neutrino_1980}
\bibinfo{author}{\bibfnamefont{R.~N.} \bibnamefont{Mohapatra}}
  \bibnamefont{and}
  \bibinfo{author}{\bibfnamefont{G.}~\bibnamefont{Senjanović}},
  \bibinfo{journal}{Physical Review Letters} \textbf{\bibinfo{volume}{44}},
  \bibinfo{pages}{912} (\bibinfo{year}{1980}).

\bibitem[{\citenamefont{Simkovic et~al.}(2007)\citenamefont{Simkovic,
  Dvornicky, and Faessler}}]{simkovic_exact_2007}
\bibinfo{author}{\bibfnamefont{F.}~\bibnamefont{Simkovic}},
  \bibinfo{author}{\bibfnamefont{R.}~\bibnamefont{Dvornicky}},
  \bibnamefont{and} \bibinfo{author}{\bibfnamefont{A.}~\bibnamefont{Faessler}},
  \bibinfo{journal}{0712.3926}  (\bibinfo{year}{2007}).

\bibitem[{\citenamefont{Johner}(1987)}]{johner_angular_1987}
\bibinfo{author}{\bibfnamefont{J.}~\bibnamefont{Johner}},
  \bibinfo{journal}{Physical Review A} \textbf{\bibinfo{volume}{36}},
  \bibinfo{pages}{1498} (\bibinfo{year}{1987}).

\bibitem[{\citenamefont{Aseev et~al.}(2000)\citenamefont{Aseev, Belesev,
  Berlev, Geraskin, Kazachenko, Kuznetsov, Lobashev, Ostroumov, Titov,
  Zadorozhny et~al.}}]{aseev_energy_2000}
\bibinfo{author}{\bibfnamefont{V.}~\bibnamefont{Aseev}},
  \bibinfo{author}{\bibfnamefont{A.}~\bibnamefont{Belesev}},
  \bibinfo{author}{\bibfnamefont{A.}~\bibnamefont{Berlev}},
  \bibinfo{author}{\bibfnamefont{E.}~\bibnamefont{Geraskin}},
  \bibinfo{author}{\bibfnamefont{O.}~\bibnamefont{Kazachenko}},
  \bibinfo{author}{\bibfnamefont{Y.}~\bibnamefont{Kuznetsov}},
  \bibinfo{author}{\bibfnamefont{V.}~\bibnamefont{Lobashev}},
  \bibinfo{author}{\bibfnamefont{R.}~\bibnamefont{Ostroumov}},
  \bibinfo{author}{\bibfnamefont{N.}~\bibnamefont{Titov}},
  \bibinfo{author}{\bibfnamefont{S.}~\bibnamefont{Zadorozhny}},
  \bibnamefont{et~al.}, \bibinfo{journal}{The European Physical Journal D -
  Atomic, Molecular, Optical and Plasma Physics} \textbf{\bibinfo{volume}{10}},
  \bibinfo{pages}{39} (\bibinfo{year}{2000}).

\bibitem[{\citenamefont{Dragoun et~al.}(2004)\citenamefont{Dragoun, Spalek, and
  Wuilleumier}}]{dragoun_increased_2004}
\bibinfo{author}{\bibfnamefont{O.}~\bibnamefont{Dragoun}},
  \bibinfo{author}{\bibfnamefont{A.}~\bibnamefont{Spalek}}, \bibnamefont{and}
  \bibinfo{author}{\bibfnamefont{F.}~\bibnamefont{Wuilleumier}},
  \bibinfo{journal}{Czechoslovak Journal of Physics}
  \textbf{\bibinfo{volume}{54}}, \bibinfo{pages}{833} (\bibinfo{year}{2004}).

\bibitem[{\citenamefont{Jonsell et~al.}(1999)\citenamefont{Jonsell, Saenz, and
  Froelich}}]{Jonsell:1999aq}
\bibinfo{author}{\bibfnamefont{S.}~\bibnamefont{Jonsell}},
  \bibinfo{author}{\bibfnamefont{A.}~\bibnamefont{Saenz}}, \bibnamefont{and}
  \bibinfo{author}{\bibfnamefont{P.}~\bibnamefont{Froelich}},
  \bibinfo{journal}{Phys. Rev.} \textbf{\bibinfo{volume}{C60}},
  \bibinfo{pages}{034601} (\bibinfo{year}{1999}).

\end{thebibliography}
   
\end{document}